\documentclass[reprint,amsmath,amssymb,aps,pra,prarmp,
prstab,
prstper,
floatfix]{revtex4-2}
\usepackage{color}
\usepackage{overpic}
\usepackage{graphicx}
\usepackage{dcolumn}
\usepackage{bm,ulem}
\usepackage[colorlinks,citecolor=blue,linkcolor=red]{hyperref}
\date{\today}
\begin{document}
\title{Overcoming the Quasi-Static Bottleneck: A Finite-Time Quantum Otto Information Engine Achieving Near-Unity Efficiency}
\author{Yang Xiao$^{1}$}
\author{Jin Wang$^{2}$}\email{jin.wang.1@stonybrook.edu}
\affiliation{ $^1\,$College of Physics, Jilin University, Changchun 130022, China\\  $^2\,$ Department of Chemistry and Department of Physics and Astronomy, State University of New York at Stony Brook, Stony Brook, New York 11794, USA}

\begin{abstract} 
Generally, quantum heat engines driven by Gibbs reservoirs achieve their maximum work and efficiency only in quasi-static processes, a constraint that hampers practical applications due to the resulting vanishing power output. To address this fundamental limitation, we propose a finite-time quantum Otto information engine (OIE) driven by a Maxwell’s demon paired with a single Gibbs reservoir. We demonstrate that the demon's measurement and feedback control can  preserve quantum coherence to extract coherent work, while  harnessing quantum internal friction as a  work source. Consequently, the OIE can produce work by only modulating the eigenstates of the Hamiltonian, and surpass the quasi-static limits of its conventional  counterpart in the  work output and the  efficiency, which  accounts for the energetic cost of the demon. Notably, the OIE can achieve near-perfect efficiency alongside positive work extraction in the rapid-driving regime.
\end{abstract}
\maketitle
\date{\today}

\section{Introduction}
Quantum heat engines \cite{JK19,RJ19,JP19,Quan07,PA19,Xiao23,kosloff06,JK17,Holubec18,Lutz24,Lutz16,DD23,Wang24,Fei22,Lutz14,man26,Kosloff14,Lutz21,ML26,Tu26,WN18} are thermodynamic devices that convert heat into  mechanical work via coupling to two reservoirs, and have been successfully implemented across various experimental platforms, including nitrogen-vacancy centers in diamond \cite{JK19}, nuclear magnetic resonance \cite{RJ19,JP19}, atomic collisions \cite{Lutz21}, photonic systems \cite{man26}, anyons \cite{ML26}, trapped single ions \cite{Lutz16,Lutz24}, and superconducting circuits \cite{Tu26}. However, a quantum heat engine typically achieves its maximum work and efficiency only at the quasi-static limit, where the operational duration tends to infinity and the power output  vanishes. This fundamental limitation severely restricts their practical applicability. To produce finite power output while maintaining high thermodynamic efficiency, various strategies have been proposed \cite{CA75, Kosloff92,Esposito10,NS16,Seifert18,VH18,Harry21,RJ19,Xiao23}, among which driving engines with non-equilibrium environments, such as squeezed reservoirs \cite{Xiao23} or effective negative-temperature reservoirs \cite{RJ19}, has emerged as a prominent route. By mitigating decoherence or converting quantum internal friction into useful work, such non-equilibrium reservoirs enable finite time engines to surpass conventional quasi-static bounds.

However, heat engines driven by non-equilibrium reservoirs face a fundamental thermodynamic challenge. Standard efficiency evaluations often neglect the external energy input required to prepare and maintain reservoirs in non-equilibrium states \cite{RJ19,GM16,JK17}. Once the actual physical cost of sustaining these non-equilibrium environments is rigorously accounted for, the global thermodynamic efficiency inevitably falls back below the Carnot limit \cite{BG15,AM15,Xiao23}. Furthermore, this dissipation is rarely quantified because ideal reservoirs are assumed to possess infinite degrees of freedom, and this cost  depends on the specific protocol used to drive the environment \cite{Lutz09,AM12}. Consequently, the engine's actual global efficiency calculation is difficult. Therefore, developing a driving mechanism that harnesses non-equilibrium effects to boost finite-time performance, while remaining self-consistent with precisely quantifiable thermodynamic costs, is a key challenge in quantum thermodynamics.

To resolve this issue,  we introduce an alternative paradigm that replaces traditional non-equilibrium reservoirs with the measurement and  control operations of a Maxwell's demon \cite{Maxwell1871,Vedral09}. As a cornerstone of non-equilibrium feedback control, Maxwell's demons can manipulate system states through measurement and feedback, thereby serving as an information-based energy source for quantum heat engines \cite{Sagawa08, Sagawa15}. Such information-driven control has been realized across platforms like Brownian systems \cite{BJ08, TA18,Sag10, GP18, YJ14}, electronics \cite{JV15, JV14, Sag14, KC17}, photonics \cite{MD16}, NMR \cite{PA16}, and superconducting circuits \cite{NC17,YM18}.
Crucially, unlike generic non-equilibrium reservoirs whose preparation costs are ill-defined or difficult to quantify, the  energy required for operating a demon, specifically the costs of information acquisition and erasure, is universally constrained and explicitly calculable via information thermodynamics \cite{Sagawa09, Landauer61}. Consequently, a demon-driven quantum engine can offer an ideal, thermodynamically self-consistent framework to exploit non-equilibrium advantages for finite time performance enhancement.

To realize this framework, we construct a finite time quantum Otto information engine (OIE) where the hot thermal reservoir is substituted by a demon's measurement and feedback operations.We show that the demon can preserve quantum coherence by weak measurement  while enabling state inversion through state-dependent feedback. Consequently, the work caused by the coherence can overcome the quantum inner friction, allowing that the finite-time OIE can extract useful work without altering the energy level gaps of the working system, relying solely on the eigenstate modulation of the Hamiltonian. Furthermore, both the  work output and the overall efficiency (which includes the demon's energetic dissipation) of the OIE, can simultaneously surpass the quasi-static limits of its conventional counterpart. Finally, the finite time OIE can deliver positive work extraction at near-perfect efficiency.
\begin{figure*}
\begin{overpic}[width=1\linewidth]{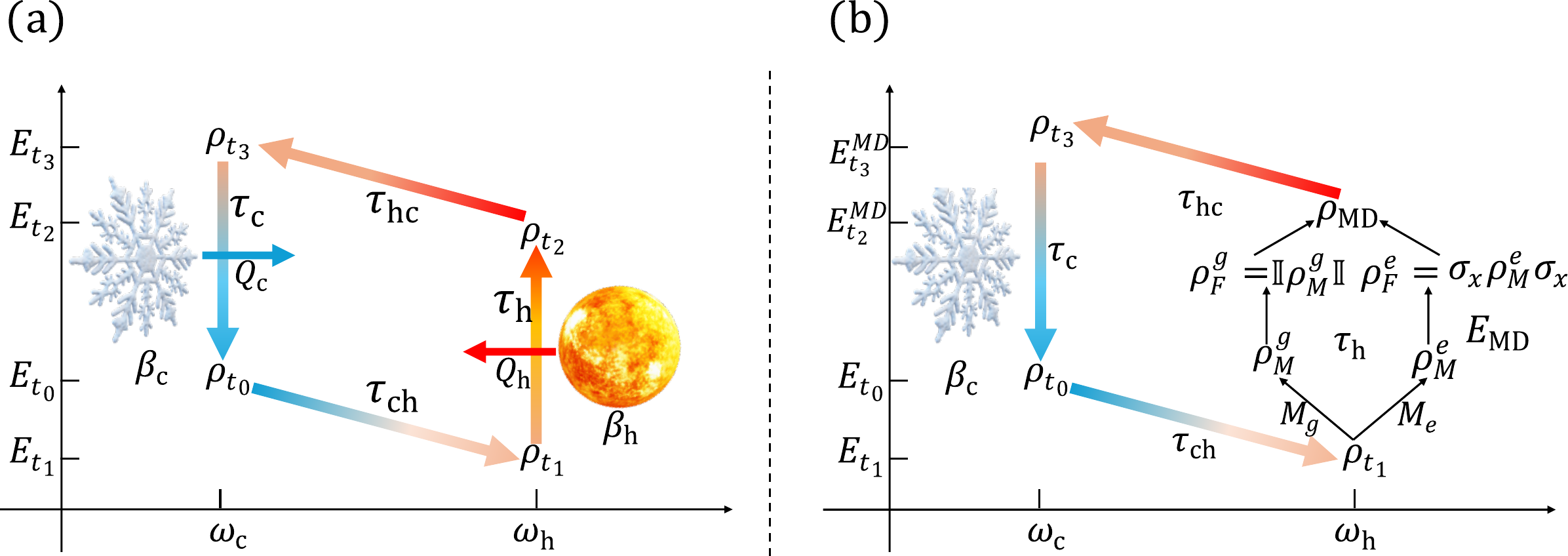}
\end{overpic}
    \caption{Schematic diagrams of the finite time quantum Otto heat engine (OHE) (a) and quantum Otto information engine (OIE) (b). (a) The conventional quantum OHE undergoes four steps: adiabatic compression ($\rho_{t_0}\to \rho_{t_1}$),   isochoric heating  ($\rho_{t_1}\to \rho_{t_2}$) at inverse temperature $\beta_\mathrm{h}$ with absorbing heat $Q_\mathrm{h}$,  adiabatic expansion ($\rho_{t_2}\to \rho_{t_3}$) and isochoric cooling ($\rho_{t_3}\to \rho_{t_0}$) at inverse temperature $\beta_\mathrm{c}$ with absorbing heat $Q_\mathrm{c}$. (b) In the quantum OIE, the isochoric heating stroke is replaced by a demon's measurement  $M_{e,g}$ followed by state-dependent feedback control, represented by the identity operator $\mathbb{I}$  and the Pauli operator $\sigma_x$.}\label{model}
\end{figure*}

\section{The limitations of the finite time quantum Otto heat engine}

The standard finite time quantum Otto heat engine (OHE) is driven by a hot reservoir at inverse temperature $\beta_\mathrm{h} = (k_B T_\mathrm{h})^{-1}$ and a cold reservoir at inverse temperature $\beta_\mathrm{c} = (k_B T_\mathrm{c})^{-1}$, and consists of two adiabatic strokes and two isochoric strokes, as shown in Fig. \ref{model}(a).
The working substance of the OHE is a spin-$1/2$ particle. The detailed dynamical cycle of the engine proceeds as follows.

\textbf{Adiabatic compression $\rho_{t_0} \to \rho_{t_1}$:}
Initially (at $t_0 = 0$), the system is in thermal equilibrium with the cold reservoir, described by the Gibbs state
$\rho_{t_0} = \exp(-\beta_\mathrm{c} H_{t_0})/\mathrm{Tr}[\exp(-\beta_\mathrm{c} H_{t_0})]$
with the initial Hamiltonian
$H_{t_0} = \sigma_x\omega_\mathrm{c}/2$ (setting $\hbar=1$). 
Here $\omega_\mathrm{c}$ is the external field frequency and $\sigma_{x,y,z}$ denote the Pauli matrices. Let $\vert{}e_{t_0}\rangle$ and $\vert{}g_{t_0}\rangle$ be the excited and ground eigenstates of $H_{t_0}$, respectively. The corresponding occupation probabilities are given by
$\rho_{t_0}^{ee} = \langle e_{t_0}|\rho_{t_0}|e_{t_0}\rangle = 1/[1+\exp(\beta_\mathrm{c}\omega_\mathrm{c})],$ and $ 
\rho_{t_0}^{gg} = \langle g_{t_0}|\rho_{t_0}|g_{t_0}\rangle = \exp(\beta_\mathrm{c}\omega_\mathrm{c})/[1+\exp(\beta_\mathrm{c}\omega_\mathrm{c})]$, respectively.
The initial internal energy of the system is then determined by 
$E_{t_0} = \mathrm{Tr}[\rho_{t_0} H_{t_0}] = \omega_\mathrm{c} \langle n_{t_0}\rangle,$
where
$\langle n_{t_0}\rangle = (\rho_{t_0}^{ee}-\rho_{t_0}^{gg})/2<0$
represents the average quantum number of the system at $t_0$.

Next, the system undergoes a unitary compression over the time interval $\tau_{\mathrm{ch}} = t_1 - t_0$. During this stroke, the energy level gap is modulated according to the time-dependent Hamiltonian
$H_{\mathrm{ch}}(t) = \omega_{\mathrm{ch}}(t)\{\cos[\pi t/(2\tau_{\mathrm{ch}})]\sigma_x + \sin[\pi t/(2\tau_{\mathrm{ch}})]\sigma_z\},$ which has been implemented experimentally \cite{JP19}
with the driving function
$\omega_{\mathrm{ch}}(t) = \omega_\mathrm{c}(1-t/\tau_{\mathrm{ch}})/2 + \omega_\mathrm{h}t/(2\tau_{\mathrm{ch}}).
$  Here, we have assumed $t_0=0$.
In the absence of heat exchange, the system dynamics is governed by the von Neumann equation \cite{HP03}
\begin{equation}\label{ms}
    \frac{d\rho_t}{dt} = -i\,[H_{\mathrm{ch}}(t),\rho_t]. 
\end{equation}
Solving Eq. (\ref{ms}), the density matrix of the system at time $t_1$ can be determined
$ \rho_{t_1} = U_{\mathrm{ch}}\,\rho_{t_0}\,U_{\mathrm{ch}}^\dagger, $
where the time-evolution operator $U_{\mathrm{ch}} = \mathcal{T}_> \exp[-i\int_{t_0}^{t_1} H_{\mathrm{ch}}(t)\,dt]$  is an unitary operator,
with $\mathcal{T}_>$ denoting the time-ordering operator.

Let $\vert{}e_{t_1}\rangle$ and $\vert{}g_{t_1}\rangle$ denote the eigenstates of the Hamiltonian $H_{t_1} = H_\mathrm{ch}(t_1) = (\omega_\mathrm{h} / 2)\,\sigma_z$. The occupation probabilities for the excited and ground states at time $t_1$ can be calculated as
$\rho_{t_1}^{ee} = \langle e_{t_1}|\rho_{t_1}|e_{t_1}\rangle
= \langle e_{t_1}|U_{\mathrm{ch}}\rho_{t_0}U_{\mathrm{ch}}^\dagger|e_{t_1}\rangle 
= \rho_{t_0}^{ee}\langle e_{t_1}|U_{\mathrm{ch}}|e_{t_0}\rangle\langle e_{t_0}|U_{\mathrm{ch}}^\dagger|e_{t_1}\rangle
 + \rho_{t_0}^{gg}\langle e_{t_1}|U_{\mathrm{ch}}|g_{t_0}\rangle\langle g_{t_0}|U_{\mathrm{ch}}^\dagger|e_{t_1}\rangle 
= \rho_{t_0}^{ee}(1-\xi) + \rho_{t_0}^{gg}\xi, 
\rho_{t_1}^{gg} = \langle g_{t_1}|\rho_{t_1}|g_{t_1}\rangle
= \langle g_{t_1}|U_{\mathrm{ch}}\rho_{t_0}U_{\mathrm{ch}}^\dagger|g_{t_1}\rangle 
= \rho_{t_0}^{ee}\langle g_{t_1}|U_{\mathrm{ch}}|e_{t_0}\rangle\langle e_{t_0}|U_{\mathrm{ch}}^\dagger|g_{t_1}\rangle
 + \rho_{t_0}^{gg}\langle g_{t_1}|U_{\mathrm{ch}}|g_{t_0}\rangle\langle g_{t_0}|U_{\mathrm{ch}}^\dagger|g_{t_1}\rangle 
= \rho_{t_0}^{ee}\xi + \rho_{t_0}^{gg}(1-\xi).$
Here
$\xi = |U_{\mathrm{ch}}^{eg}|^2
    = |U_{\mathrm{ch}}^{ge}|^2$
represents the transition probability between the two energy levels during this stroke \cite{RJ19,JP19}.
The probability amplitudes $U_{\mathrm{ch}}^{eg} = \langle e_{t_1}|U_{\mathrm{ch}}|g_{t_0}\rangle$ and $U_{\mathrm{ch}}^{ge} = \langle g_{t_1}|U_{\mathrm{ch}}|e_{t_0}\rangle$.
Defining $U_{\mathrm{ch}}^{ee} = \langle e_{t_1}|U_{\mathrm{ch}}|e_{t_0}\rangle$ and $U_{\mathrm{ch}}^{gg} = \langle g_{t_1}|U_{\mathrm{ch}}|g_{t_0}\rangle$, the unitarity condition $U_{\mathrm{ch}}U_{\mathrm{ch}}^\dagger = \mathbb{I}$ leads to 
$U_{\mathrm{ch}}^{ee}U_{\mathrm{ch}}^{ge*} + U_{\mathrm{ch}}^{eg}U_{\mathrm{ch}}^{gg*} = \langle e_{t_1}|U_{\mathrm{ch}}U_{\mathrm{ch}}^\dagger|g_{t_1}\rangle = 0$, which establishes the relation
$U_{\mathrm{ch}}^{ee}U_{\mathrm{ch}}^{ge*} = -U_{\mathrm{ch}}^{eg}U_{\mathrm{ch}}^{gg*}$.
Consequently, the off-diagonal elements of the density matrix $\rho_{t_1}$ are given by
$\rho_{t_1}^{eg} = \langle e_{t_1}|\rho_{t_1}|g_{t_1}\rangle
= \langle e_{t_1}|U_{\mathrm{ch}}\rho_{t_0}U_{\mathrm{ch}}^\dagger|g_{t_1}\rangle = \rho_{t_0}^{ee}\,U_{\mathrm{ch}}^{ee}U_{\mathrm{ch}}^{ge*}
 + \rho_{t_0}^{gg}\,U_{\mathrm{ch}}^{eg}U_{\mathrm{ch}}^{gg*} = 2\,U_{\mathrm{ch}}^{ee}U_{\mathrm{ch}}^{ge*}\,\langle n_{t_0}\rangle$,
and $\rho_{t_1}^{ge} = (\rho_{t_1}^{eg})^* = 2\,U_{\mathrm{ch}}^{ee*}U_{\mathrm{ch}}^{ge}\langle n_{t_0}\rangle$. Following Ref. \cite{TB14}, the quantum coherence of the system at time $t_1$ is quantified by the $l_1$-norm of coherence as
\begin{equation}
C(\rho_{t_1})=|\rho_{t_1}^{eg}|+|\rho_{t_1}^{ge}|=-2|U_{\mathrm{ch}}^{ee}U_{\mathrm{ch}}^{ge*}|\langle n_{t_0}\rangle.
\end{equation}

Finally, the internal energy of the system at $t_1$ reads
$E_{t_1} = \mathrm{Tr}[\rho_{t_1} H_{t_1}]
       = \omega_\mathrm{h}(\rho_{t_1}^{ee}-\rho_{t_1}^{gg})/2
       = \omega_\mathrm{h} \langle n_{t_1}\rangle,$
where $\langle n_{t_1}\rangle = (\rho_{t_1}^{ee}-\rho_{t_1}^{gg})/2=(1-2\xi)\langle n_{t_0}\rangle$ denotes the corresponding average quantum number. Because this stroke is  adiabatic with zero heat exchange, the work performed by the system during compression is given by $W_{\mathrm{ch}} = E_{t_0} - E_{t_1}.$

\textbf{Isochoric heating $\rho_{t_1} \to \rho_{t_2}$:}
During this stroke, the system weakly interacts with the hot reservoir for a duration $\tau_\mathrm{h} = t_2 - t_1$ under the fixed Hamiltonian $H_{t_1}$. The dynamical evolution of the system is governed by the Lindblad master equation \cite{HP03}
\begin{eqnarray}\label{me}
   \frac{d\rho_t}{dt} &=& -i[H_\mathrm{h}, \rho_t] 
+ \gamma_{\downarrow}^\mathrm{h}\Bigl(\sigma_-\rho_t\sigma_+ - \frac{1}{2}\{\sigma_+\sigma_-,\rho_t\}\Bigr)\nonumber\\
&+& \gamma_{\uparrow}^\mathrm{h}\Bigl(\sigma_+\rho_t\sigma_- - \frac{1}{2}\{\sigma_-\sigma_+,\rho_t\}\Bigr), 
\end{eqnarray}
where the decay and excitation rates are $\gamma_{\downarrow}^\mathrm{h} = \gamma_0(1+N_\mathrm{h})$ and $\gamma_{\uparrow}^\mathrm{h} = \gamma_0 N_\mathrm{h}$, respectively. Here, $N_\mathrm{h} = (e^{\beta_\mathrm{h}\omega_\mathrm{h}}-1)^{-1}$ is the average thermal photon number of the hot reservoir, and $\gamma_0$ represents the spontaneous emission rate.

By solving Eq. (\ref{me}),   the state of the system  at time $t_2$ can  be determined $\rho_{t_2}=\sum_{j_{t_1},j_{t_1}^{'}}\rho_{t_2}^{j_{t_1},j_{t_1}^{'}}|j_{t_1}\rangle\langle j_{t_1}^{'}|$ $(j_{t_1},j_{t_1}^{'}=e,g)$
with the matrix elements given by
$\rho_{t_2}^{ee} = p_\mathrm{h}^{\mathrm{eq}} + \bigl(\rho_{t_1}^{ee} - p_\mathrm{h}^{\mathrm{eq}}\bigr) e^{-\Gamma_\mathrm{h} \tau_\mathrm{h}}, 
\rho_{t_2}^{eg} = \rho_{t_1}^{eg} e^{-\Gamma_\mathrm{h}\tau_\mathrm{h}/2} e^{-i\omega_\mathrm{h} \tau_\mathrm{h}},
\rho_{t_2}^{ge} = \rho_{t_2}^{eg*}, 
\rho_{t_2}^{gg} =1 - p_\mathrm{h}^{\mathrm{eq}} + \bigl(p_\mathrm{h}^{\mathrm{eq}} - \rho_{t_1}^{ee}\bigr) e^{-\Gamma_\mathrm{h} \tau_\mathrm{h}}.$
Here $p_\mathrm{h}^{\mathrm{eq}} = 1/(1+e^{\beta_\mathrm{h}\omega_\mathrm{h}})$ is the excited-state probability of the system   at thermal equilibrium with the hot reservoir,
and $\Gamma_\mathrm{h} = \gamma_{\downarrow}^\mathrm{h} + \gamma_{\uparrow}^\mathrm{h}$  denotes the total relaxation rate.
Accordingly, the quantum coherence of the system at time $t_2$  can be obtained 
$C(\rho_{t_2})=|\rho_{t_2}^{eg}|+|\rho_{t_2}^{ge}|=C(\rho_{t_1})e^{-\Gamma_\mathrm{h} \tau_\mathrm{h}/2}$ which exhibits an exponential decay with respect to $\Gamma_\mathrm{h} \tau_\mathrm{h}$.

Based on $\rho_{t_2}$, the internal energy of the system at time $t_2$ is obtained as
$E_{t_2} = \mathrm{Tr}[\rho_{t_2} H_{t_1}] = \omega_\mathrm{h} \langle n_{t_2}\rangle,$
where the average quantum number  at $t_2$ reads 
$\langle n_{t_2}\rangle = (\rho_{t_2}^{ee} - \rho_{t_2}^{gg})/2
= \langle n_\mathrm{h}^{\mathrm{eq}}\rangle + (\langle n_{t_1}\rangle - \langle n_\mathrm{h}^{\mathrm{eq}}\rangle) e^{-\Gamma_\mathrm{h} \tau_\mathrm{h}},$
with $\langle n_\mathrm{h}^{\mathrm{eq}}\rangle = (2p_\mathrm{h}^{\mathrm{eq}}-1)/2$ representing the thermal equilibrium average quantum number.
Furthermore, since no work is done during this isochoric stroke, the heat absorbed from the hot reservoir is given by
\begin{equation}\label{qh}
    Q_\mathrm{h} = E_{t_2} - E_{t_1}
= \omega_\mathrm{h}\bigl[\langle n_{t_2}\rangle - (1-2\xi)\langle n_{t_0}\rangle\bigr], 
\end{equation}

\textbf{Adiabatic expansion $\rho_{t_2} \to \rho_{t_3}$:}
During this stroke, the system is again isolated from the thermal reservoirs. The Hamiltonian is modulated back to its initial configuration $H_{t_0}$ over a duration $\tau_\mathrm{hc} = t_3 - t_2$.
The time-dependent driving protocol is defined as
$H_{\mathrm{hc}}(t)=H_\mathrm{ch}(t_1+t_2-t)$, which guarantees that this stroke is the time-reversed counterpart of the adiabatic compression, yielding $\tau_\mathrm{hc}=\tau_\mathrm{ch}=\tau$.

Following a procedure analogous to the compression stroke, the density matrix of the system at time $t_3$ is given by
$\rho_{t_3} = U_{\mathrm{hc}}\,\rho_{t_2}\,U_{\mathrm{hc}}^\dagger$,
with the evolution operator
$U_{\mathrm{hc}} = \mathcal{T}_> \exp[-i\int_{t_2}^{t_3} H_{\mathrm{hc}}(t)\,dt]$.
The corresponding matrix elements of $\rho_{t_3}$ can be expressed as $\rho_{t_3}^{ee}=\rho_{t_2}^{ee}(1-\xi) + \rho_{t_2}^{gg}\xi-2\mathrm{Re}(\rho_{t_2}^{ge}U_{hc}^{gg}U_{hc}^{ge*})$,
$\rho_{t_3}^{eg}=2\langle n_{t_2}\rangle+\rho_{t_2}^{eg}U_\mathrm{hc}^{ee}U_\mathrm{hc}^{gg*}+\rho_{t_2}^{ge}U_\mathrm{hc}^{eg}U_\mathrm{hc}^{ge*}$,$\rho_{t_3}^{ge}=\rho_{t_3}^{eg*}$, and $\rho_{t_3}^{gg}=\rho_{t_2}^{ee}\xi+\rho_{t_2}^{gg}(1-\xi)+2\mathrm{Re}(\rho_{t_2}^{ge}U_{hc}^{gg}U_{hc}^{ge*})$ (see Appendix \ref{app1}). Here, the probability amplitudes $U_\mathrm{hc}^{kj}=\langle k_{t_0}|U_\mathrm{hc}|j_{t_1}\rangle$ $(k,j=e,g)$, and $\xi=|U_\mathrm{hc}^{j\neq k}|^2 =|U_\mathrm{hc}^{j\neq k}|^2$ \cite{JP19,RJ19}.
The internal energy of the system at $t_3$ can then be written as
$E_{t_3} = \mathrm{Tr}[\rho_{t_3} H_{t_0}]=\omega_\mathrm{c}\langle n_{t_3}\rangle$,
where the average quantum number $\langle n_{t_3}\rangle=\langle n_{t_2}\rangle(1-2\xi)-2\zeta_\mathrm{hc}$, with 
$\zeta_\mathrm{hc}=\mathrm{Re}(\rho_{t_2}^{ge}U_\mathrm{hc}^{gg}U_\mathrm{hc}^{ge*})$
represent the quantum number change caused by the coherence. 
Consequently, the work extracted by the system during this adiabatic expansion stroke is
$W_{\mathrm{hc}} = E_{t_2} - E_{t_3}$.

\textbf{Isochoric cooling $\rho_{t_3} \to \rho_{t_0}$:}
During this final stroke, the system Hamiltonian remains fixed at $H_{t_0}$, and the system is coupled to the cold reservoir for a time duration $\tau_\mathrm{c}$.
 Following the master equation approach described in the heating stroke, the density matrix of the system at time $t_\mathrm{cyc}=t_3+\tau_\mathrm{c}$   can be determined 
$\rho_{t_\mathrm{cyc}} =\sum_{k_{t_0},k_{t_0}^{'}}\rho_{t_\mathrm{cyc}}^{k_{t_0},k_{t_0}^{'}}|k_{t_0}\rangle\langle k_{t_0}^{'}|$ $(k_{t_0},k_{t_0}^{'}=e,g)$
with the matrix elements given by
$\rho_{t_\mathrm{cyc}}^{ee} = p_\mathrm{c}^{\mathrm{eq}} + \bigl(\rho_{t_3}^{ee} - p_\mathrm{c}^{\mathrm{eq}}\bigr) e^{-\Gamma_\mathrm{c} \tau_\mathrm{c}}, 
\rho_{t_\mathrm{cyc}}^{eg} = \rho_{t_3}^{eg} e^{-\Gamma_\mathrm{c} \tau_\mathrm{c}/2} \, e^{-i\omega_\mathrm{c} \tau_\mathrm{c}},
\rho_{t_\mathrm{cyc}}^{ge} =\rho_{t_\mathrm{cyc}}^{eg*}, 
\rho_{t_\mathrm{cyc}}^{gg} = 1 - p_\mathrm{c}^{\mathrm{eq}} + \bigl(p_\mathrm{c}^{\mathrm{eq}} - \rho_{t_3}^{ee}\bigr) e^{-\Gamma_\mathrm{c} \tau_\mathrm{c}}.$
Here, $p_\mathrm{c}^{\mathrm{eq}} = (1+e^{\beta_\mathrm{c}\omega_\mathrm{c}})^{-1}$ is the excited-state population at thermal equilibrium with the cold reservoir, and $\Gamma_c = \gamma_{\downarrow}^c + \gamma_{\uparrow}^c$ is the total transition rate with $\gamma_{\downarrow}^\mathrm{c} = \gamma_0(1+N_\mathrm{c})$ and $\gamma_{\uparrow}^\mathrm{c} = \gamma_0 N_\mathrm{c}$. The quantity $N_\mathrm{c} = (e^{\beta_\mathrm{c}\omega_\mathrm{c}}-1)^{-1}$ denotes the average thermal photon number of the cold reservoir.
In the long-time limit ($\Gamma_\mathrm{c}\tau_\mathrm{c} \gg 1$), the system fully relaxes back to the initial Gibbs state $\rho_{t_0}$, thereby completing a closed cyclic operation. Since no mechanical work is exchanged during this isochoric stroke, the heat absorbed from the cold reservoir is given by
$Q_\mathrm{c}= E_{t_0} - E_{t_3}$.

After one complete cycle, the total work output $W_{\mathrm{fin}}^{\mathrm{Otto}} = W_{\mathrm{ch}} + W_{\mathrm{hc}}$ of the finite time quantum OHE can be decomposed into three distinct physical contributions \cite{Xiao23}
\begin{eqnarray}
    W_{\mathrm{fin}}^{\mathrm{OHE}}=
    W_\mathrm{trls}^{\mathrm{OHE}}+W_\mathrm{fri}^{\mathrm{OHE}}+W_\mathrm{coh}^{\mathrm{OHE}},
\end{eqnarray}
where $W_\mathrm{trls}^{\mathrm{OHE}}=(\omega_\mathrm{h} - \omega_\mathrm{c})(\langle n_{t_2}\rangle - \langle n_{t_0}\rangle)$ denotes the work output in the transitionless case,  $W_\mathrm{fri}^{\mathrm{OHE}}=2\xi(\omega_\mathrm{h}\langle n_{t_0}\rangle + \omega_\mathrm{c}\langle n_{t_2}\rangle)$ represents the work associated with quantum internal friction.
Because both $\langle n_{t_0}\rangle$ and $\langle n_{t_2}\rangle$ are negative, this friction term  diminishes the  work output.
Finally, the third term $W_\mathrm{coh}^{\mathrm{OHE}}=2\omega_\mathrm{c}\zeta_{\mathrm{hc}}$ captures the work contribution from quantum coherence.
When $\zeta_{\mathrm{hc}} > 0$, quantum coherence serves as an additional energetic resource that enhances work extraction.
\begin{figure*}
    \centering
    \includegraphics[width=1\linewidth]{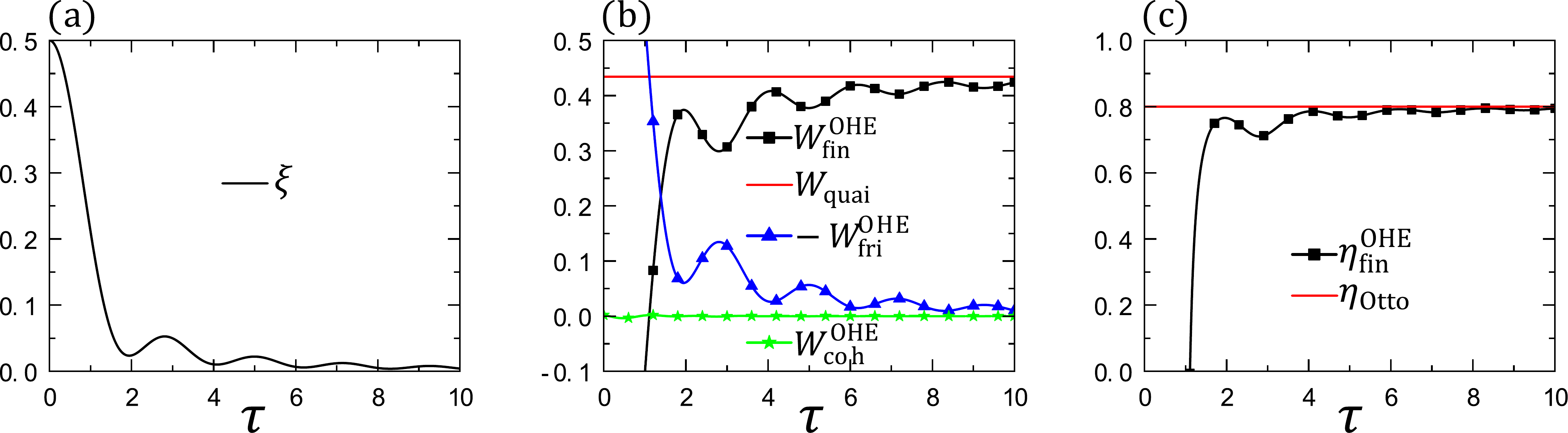}
    \caption{(a) The transition probability $\xi$, along with (b) the work output $W$ and (c) the efficiency $\eta$ of the finite-time quantum OHE as functions of the driving time $\tau$. The chosen simulation parameters are $\gamma_0 = 10,\beta_\mathrm{c}=10\beta_\mathrm{h} = 1, \omega_\mathrm{h}=5\omega_\mathrm{c} =5,$ and $\tau_\mathrm{c} = 10\tau_\mathrm{h}=2.$ Under this parameter regime, the damping factor evaluates to $e^{-\Gamma_\mathrm{c}\tau_\mathrm{c}} \approx 1.6 \times 10^{-19}$, which ensures that the working system fully relaxes back to its initial Gibbs state at the end of each cycle, thereby maintaining a strictly closed cyclic operation.}
    \label{Otto}
\end{figure*}
From the definition of $\zeta_{\mathrm{hc}}$, we obtain the bound $ \zeta_{\mathrm{hc}} \le |\rho_{t_2}^{ge}||U_{\mathrm{hc}}^{gg}||U_{\mathrm{hc}}^{ge*}|=C(\rho_{t_2})|U_{\mathrm{hc}}^{gg}||U_{\mathrm{hc}}^{ge*}|/2
= -2(1-\xi)\xi e^{-\Gamma_\mathrm{h} \tau_\mathrm{h}/2}\,\langle n_{t_0}\rangle,$  
which yields the following  inequality
$2\xi(\omega_\mathrm{h}\langle n_{t_0}\rangle + \omega_\mathrm{c}\langle n_{t_2}\rangle) + 2\omega_\mathrm{c}\zeta_{\mathrm{hc}}
 < 2\omega_c[\xi(\langle n_{t_0}\rangle + \langle n_{t_2}\rangle) + \zeta_{\mathrm{hc}}] 
 \le 2\omega_\mathrm{c}\,\xi[\langle n_{t_0}\rangle + \langle n_{t_2}\rangle - 2(1-\xi)e^{-\Gamma_\mathrm{h} \tau_\mathrm{h}/2}\,\langle n_{t_0}\rangle]  = 2\omega_\mathrm{c}\,\xi[\langle n_\mathrm{h}^{\mathrm{eq}}\rangle(1-e^{-\Gamma_\mathrm{h} \tau_\mathrm{h}}) + \langle n_{t_0}\rangle F(k_\mathrm{h})].$
Here we  have defined $k_\mathrm{h} = e^{-\Gamma_\mathrm{h} \tau_\mathrm{h}/2}$ with $0 < k_\mathrm{h} < 1$, and introduced the quadratic polynomial
$F(k_\mathrm{h}) = (1-2\xi)k_\mathrm{h}^2 - 2(1-\xi)k_\mathrm{h} + 1.$
The symmetry axis of $F(k_\mathrm{h})$ lies at $k_\mathrm{h} = (1-\xi)/(1-2\xi)$, which is strictly greater than $1$ for any non-adiabatic transition probability $\xi < 1/2$ \cite{RJ19}, as shown in Fig. \ref{Otto}(a).
Consequently, $F(k_\mathrm{h}) > F(1) = 0$ holds for all $0 < k_\mathrm{h} < 1$. This implies that
 $ \xi(\langle n_{t_0}\rangle + \langle n_{t_2}\rangle) + \zeta_{\mathrm{hc}} < 0,$
which  establishes the relation $W_\mathrm{fri}^\mathrm{OHE}+W_\mathrm{coh}^\mathrm{OHE}<0$. Thus, we have
\begin{eqnarray}\label{WOtto}
W_{\mathrm{fin}}^{\mathrm{OHE}} &<& (\omega_\mathrm{h} - \omega_\mathrm{c})(\langle n_{t_2}\rangle - \langle n_{t_0}\rangle)\nonumber\\
&< &W_{\mathrm{qs}}^{\mathrm{OHE}} = (\omega_\mathrm{h} - \omega_\mathrm{c})(\langle n_\mathrm{h}^{\mathrm{eq}}\rangle - \langle n_{t_0}\rangle).   
\end{eqnarray}
where $W_{\mathrm{qs}}^{\mathrm{OHE}}$ denotes the work output of the OHE under quasi-static conditions \cite{Quan07}. 

The efficiency of the finite time quantum OHE is defined as the net work output divided by the heat absorbed from the hot reservoir:
\begin{eqnarray}\label{eff}
\eta_{\mathrm{fin}}^{\mathrm{OHE}}&=& \frac{W_{\mathrm{fin}}^{\mathrm{OHE}}}{Q_\mathrm{h}}\nonumber\\
&=&\eta_{\mathrm{Otto}} + \frac{2\omega_\mathrm{c}[\xi(\langle n_{t_0}\rangle + \langle n_{t_2}\rangle) + \zeta_{\mathrm{hc}}]}{Q_\mathrm{h}}\nonumber\\
&\leq& \eta_{\mathrm{Otto}}
\end{eqnarray}
where $\eta_{\mathrm{Otto}} = 1 - \omega_\mathrm{c}/\omega_\mathrm{h}$ represents the quasi-static Otto efficiency \cite{Quan07}. Applying inequality $ \xi(\langle n_{t_0}\rangle + \langle n_{t_2}\rangle) + \zeta_{\mathrm{hc}} < 0$  once more, we strictly obtain $\eta_{\mathrm{fin}}^{\mathrm{OHE}} < \eta_{\mathrm{Otto}}$. Although this inequality was mentioned in Ref.~\cite{PA19}, no explicit proof was provided, nor was the upper bound on work output in Eq.~(\ref{WOtto}) established.

Together, Eqs.~(\ref{WOtto}) and (\ref{eff}) rigorously prove that both the work output and the efficiency of the finite time quantum OHE driven by thermal Gibbs reservoirs are strictly lower than their corresponding values under quasi-static conditions, as illustrated in Figs.~\ref{Otto}(b) and (c). The underlying physical origin lies in that due to the existence of the non-equilibrium factor $k_\mathrm{h}$, the the maximum average quantum number  transfer induced by the quantum coherence can never exceed the average quantum number deficit caused by the quantum internal friction. Consequently, the positive work contribution from quantum coherence is insufficient to compensate for the negative work generated by quantum internal friction, as shown in Fig.~\ref{Otto}(b).

However, these quasi-static performance bottlenecks can be completely resolved by incorporating a Maxwell's demon, which can preserve quantum coherence to boost the coherent work while effectively converting quantum internal friction into useful work output, as we shall discuss in detail below.

\section{The finite time quantum Otto information engine}
\subsection{The performance of the  finite time quantum Otto information engine}
As illustrated in Fig.~\ref{model}(b), the detailed operational cycle of the finite-time quantum Otto information engine (OIE) proceeds through the following four steps.

First, the adiabatic compression stroke of the OIE ($\rho_{t_0} \to \rho_{t_1}$) is identical to that of the conventional finite-time OHE described above. 

Then, the conventional hot thermal reservoir is replaced by the measurement and state-dependent feedback operations of a Maxwell's demon. 
Immediately following the completion of the adiabatic compression stroke, the demon performs a generalized measurement on $\rho_{t_1}$.
The corresponding measurement operators are given by
\begin{equation}
\begin{aligned}
M_e &= \sqrt{1-\epsilon}\;|e_{t_1}\rangle\langle e_{t_1}| + \sqrt{\epsilon}\;|g_{t_1}\rangle\langle g_{t_1}|,\\
M_g &= \sqrt{\epsilon}\;|e_{t_1}\rangle\langle e_{t_1}| + \sqrt{1-\epsilon}\;|g_{t_1}\rangle\langle g_{t_1}|,
\end{aligned}
\end{equation}
where  $\epsilon \in [0, 1]$.
These operators satisfy the completeness relation $\sum_m M_m^\dagger M_m = \mathbb{I}$.

Upon obtaining outcome $m = e$, the post-measurement state becomes $\rho_M^e=M_e \rho_{t_1} M_e^\dagger/p_e$  with probability $p_e =\mathrm{Tr}(M_e \rho_{t_1} M_e^\dagger) =\rho_{t_1}^{ee}(1-\epsilon) + \rho_{t_1}^{gg}\epsilon$.
Similarly, for outcome $m = g$, the post-measurement state reads 
$\rho_M^g =M_g \rho_{t_1} M_g^\dagger/p_g$
with probability $p_g =\mathrm{Tr}(M_g \rho_{t_1} M_g^\dagger)=\rho_{t_1}^{ee}\epsilon + \rho_{t_1}^{gg}(1-\epsilon)$.
By combining $p_e$  and $p_g$, the physical meaning of $\epsilon$ can be identified: it represents the measurement error of the demon. When $\epsilon=0$ , the demon's measurement is perfectly accurate $p_{e,g}=\rho_{t_1}^{ee,gg}$. As $\epsilon$ increases, the measurement outcome gradually deviates from the  value of the corresponding measurement operator, and when $\epsilon=1$, the demon's measurement is completely erroneous $p_{e,g}=\rho_{t_1}^{gg,ee}$.

The ensemble-averaged internal energy of the system  after the measurement is calculated as
$E_M = p_e \mathrm{Tr}[\rho_M^e H_{t_1}] + p_g \mathrm{Tr}[\rho_M^g H_{t_1}] = \omega_\mathrm{h}\langle n_{t_1}\rangle = E_{t_1},$ which
demonstrates that the measurement stroke itself is non-invasive in energy, meaning no net average work or energy is directly injected into or extracted from the working system during the measurement process.

\begin{figure}
    \centering
    \includegraphics[width=0.9\linewidth]{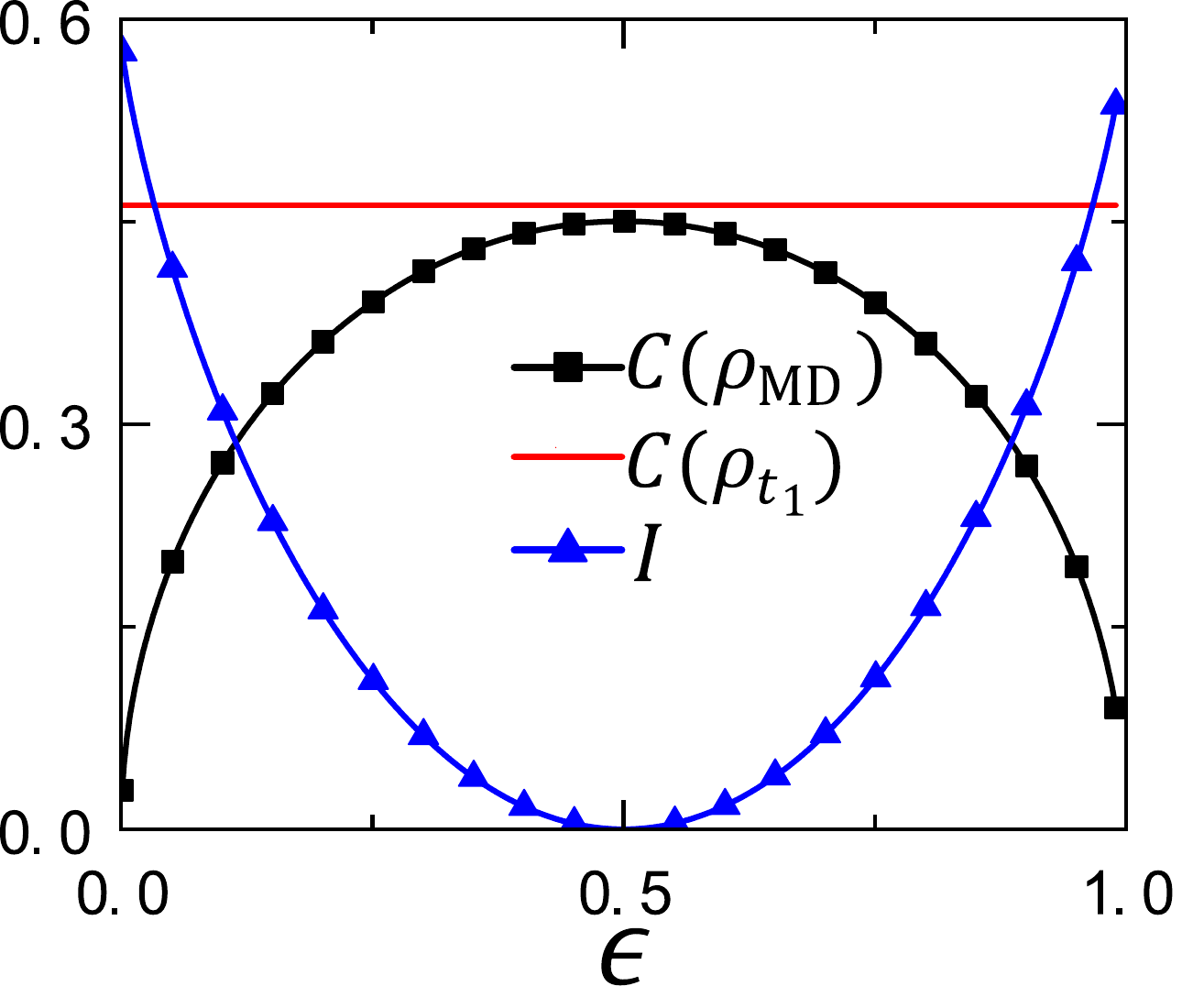}
    \caption{Quantum coherence of the OIE at times $t_1$ and $t_2$, along with the acquired quantum mutual information $I$, as functions of the measurement error $\epsilon$ for driving time $\tau = 0.1$. The other parameters are the same Fig. \ref{Otto}.}
    \label{coh}
\end{figure}

According the definition of the quantum mutual information \cite{Sagawa08},  the  information acquired by the demon through this measurement is
\begin{equation}
    I = S(\rho_{t_1}) - p_e S(\rho_M^e) - p_g S(\rho_M^g),
\end{equation}
where $S(\rho) = -\mathrm{Tr}(\rho\ln\rho)$ represents the von Neumann entropy. This acquired mutual information $I$  decreases and then increases as the measurement error $\epsilon$ increases. In the projective limit ($\epsilon = 0,1$), the measurement reduces to a standard selective measurement, and $I$ reaches its maximum value $-p_e \ln p_e-p_g \ln p_g$, which is the Shannon entropy of state $\rho_{t_1}$, as depicted in Fig.~\ref{coh}. Conversely, at $\epsilon = 0.5$, the measurement degenerate into an uninformative weak measurement, yielding $I = 0$.
Furthermore, based on the  bounds of the information thermodynamics \cite{Sagawa09}, the minimal thermodynamic energy required for the demon to perform the measurement and  reset its memory in an environment at inverse temperature $\beta_\mathrm{c}$ is given by $I / \beta_c$.

Following the measurement, the demon applies a state-dependent feedback operation conditioned on the outcome $m$: If $m=e$, the demon flips the system energy levels,
    $\rho_F^e = \sigma_x \rho_M^e \sigma_x$.
    If $m=g$, the demon does nothing, $\rho_F^g = \mathbb{I}\rho_M^g\mathbb{I}$. Here, $\mathbb{I}$ represents the identity operator.
Averaging over all possible measurement outcomes, the ensemble-averaged density matrix $\rho_{\mathrm{MD}} = p_e \rho_F^e + p_g \rho_F^g$ after feedback control takes the  matrix form in the basis $\{\vert{}e_{t_1}\rangle, \vert{}g_{t_1}\rangle\}$
\begin{equation}\label{rhomd}
   \rho_{\mathrm{MD}} 
= \begin{pmatrix}
\epsilon & 2\mathrm{Re}[\rho_{t_1}^{eg}] \sqrt{(1-\epsilon)\epsilon} \\
2\mathrm{Re}[\rho_{t_1}^{eg}] \sqrt{(1-\epsilon)\epsilon} & 1-\epsilon
\end{pmatrix}. 
\end{equation}
Assuming the duration of the demon's measurement-and-feedback process is $\tau_\mathrm{h}$, Eq. (\ref{rhomd}) indicates that the  coherence of the state $\rho_\mathrm{MD}$
\begin{equation}
    C(\rho_\mathrm{MD})=4|\mathrm{Re}[\rho_{t_1}^{eg}]| \sqrt{(1-\epsilon)\epsilon}
\end{equation}
 is governed exclusively by the measurement error $\epsilon$, entirely bypassing the exponential coherence decay induced by thermal reservoir couplings during conventional driving. This coherence increases in the regime $0\leq \epsilon \leq 0.5$  and decreases in the regime $0.5\leq \epsilon \leq 1$ as  the measurement error $\epsilon$ increases, a as shown in Fig. \ref{coh}. In the projective measurement limits ($\epsilon = 0$ or $\epsilon = 1$), the demon completely destroys quantum coherence $    C(\rho_\mathrm{MD})=0$.
Conversely, at $\epsilon = 0.5$ (the limit of the weak measurement), the real part of the off-diagonal elements of $\rho_{t_1}$ is perfectly preserved, yielding a maximum residual coherence  $2|\mathrm{Re}[\rho_{t_1}^{eg}]|$.  Provided that the imaginary part of $\rho_{t_1}^{eg}$ is negligible relative to its real part ($\vert{}\mathrm{Im}[\rho_{t_1}^{eg}]\vert{} \ll \vert{}\mathrm{Re}[\rho_{t_1}^{eg}]\vert{}$), the quantum coherence generated during the first adiabatic stroke is nearly intact after the demon's action (see Fig.~\ref{coh}).

Based on Eq. (\ref{rhomd}),  the average internal energy after the demon's feedback is 
$E_{t_2}^{\mathrm{MD}} = \mathrm{Tr}[\rho_{\mathrm{MD}} H_{t_1}] = \omega_\mathrm{h} \langle n_{t_2}^{\mathrm{MD}}\rangle,$
where $\langle n_{t_2}^{\mathrm{MD}}\rangle = \epsilon - 1/2$ denotes the average quantum number of the system at time $t_2$ in the OIE, which increases monotonically as the measurement error $\epsilon$ increases. 
Consequently, 
the energy $E_{\mathrm{MD}} = E_{t_2}^{\mathrm{MD}} - E_{t_1}$ injected into the system by the demon during this feedback stroke is given by
\begin{equation}\label{Emd}
  E_{\mathrm{MD}}=\omega_\mathrm{h}[\epsilon-\frac{1}{2}-(1-2\xi)\langle n_{t_0}\rangle].
\end{equation}

Next, the system undergoes an adiabatic expansion stroke from $H_{t_1}$ to $H_{t_0}$ over a duration $\tau$, governed by the same unitary evolution mechanism as the conventional OHE. 
Using the same method that gave the internal energy at the end of the OHE's expansion stroke,  the internal energy of the quantum OIE at time $t_3$ can be written as 
$E_{t_3}^{\mathrm{MD}}
= \omega_\mathrm{c} \langle n_{t_3}^{\mathrm{MD}}\rangle,$
where the post-expansion average quantum number is given by $\langle n_{t_3}^{\mathrm{MD}}\rangle = \langle n_{t_2}^{\mathrm{MD}}\rangle (1-2\xi) - 2\zeta_{\mathrm{hc}}^{\mathrm{MD}}$. Here, the coherence-induced quantum number shift term is expressed as 
$\zeta_{\mathrm{hc}}^{\mathrm{MD}} = \mathrm{Re}[\rho_{\mathrm{MD}}^{ge}\,U_{\mathrm{hc}}^{gg}\,U_{\mathrm{hc}}^{ge*}]$,
with the off-diagonal matrix element after feedback control being $\rho_{\mathrm{MD}}^{ge}=2\mathrm{Re}[\rho_{t_1}^{eg}] \sqrt{(1-\epsilon)\epsilon}$.
Accordingly, the  work extracted during this adiabatic expansion stroke is
$W_{\mathrm{hc}}^{\mathrm{OIE}} = E_{t_2}^{\mathrm{MD}} - E_{t_3}^{\mathrm{MD}}.$

Finally, let the system return to its initial state within $\tau_\mathrm{c}$ time according to the method in the OHE, completing a cycle.

\begin{figure*}
    \centering
    \includegraphics[width=1\linewidth]{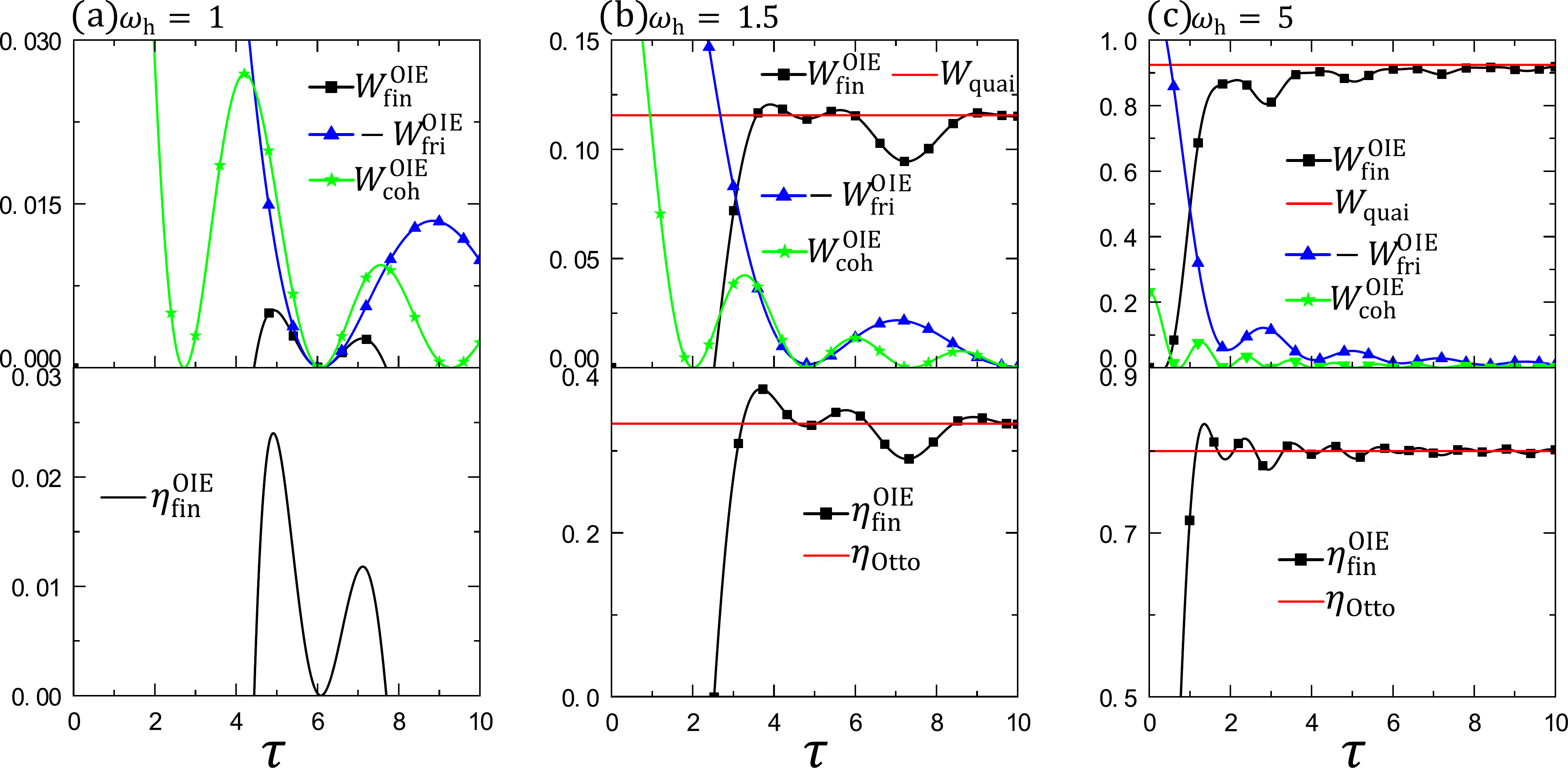}
    \caption{The output work and the efficiency of the finite time quantum OIE as functions of the driving time $\tau$ with given $\omega_\mathrm{h}$ under measurement error $\epsilon=0.5$. The other parameters are the same as Fig. \ref{Otto}.}
    \label{e0.5}
\end{figure*}

After a cycle, the  work output $W_{\mathrm{fin}}^{\mathrm{OIE}}= W_{\mathrm{ch}} + W_{\mathrm{hc}}^{\mathrm{OIE}}$ of the OIE has the same form as the work $W_{\mathrm{fin}}^{\mathrm{OHE}}$, and reads
\begin{equation}
W_{\mathrm{fin}}^{\mathrm{OIE}}= W_\mathrm{trls}^\mathrm{OIE}+W_\mathrm{fri}^\mathrm{OIE}+W_\mathrm{coh}^\mathrm{OIE}
\end{equation}
where $W_\mathrm{trls}^\mathrm{OIE}=(\omega_\mathrm{h} - \omega_\mathrm{c})(\langle n_{t_2}^{\mathrm{MD}}\rangle - \langle n_{t_0}\rangle)$,   $W_\mathrm{fri}^\mathrm{OIE}=2\xi(\omega_\mathrm{h}\langle n_{t_0}\rangle + \omega_\mathrm{c}\langle n_{t_2}^{\mathrm{MD}}\rangle)$, and $W_\mathrm{coh}^\mathrm{OIE}=2\omega_\mathrm{c}\zeta_{\mathrm{hc}}^{\mathrm{MD}}$.
Here, $W_{\mathrm{trls}}^{\mathrm{OIE}}$ corresponds to the work produced by the OIE in the quasi-static limit.

To establish a clear physical mapping between the  OIE and the OHE driven by the conventional thermal reservoir, we define an effective inverse temperature $\beta_\mathrm{h}^{\mathrm{eff}}$ for the state after the demon's action via the detailed balance relation $\exp(\beta_\mathrm{h}^\mathrm{eff}\omega_\mathrm{h})=(1-\epsilon)/\epsilon$. Under this definition, an OIE operating with measurement error $\epsilon$ is mathematically equivalent to a OHE driven by a hot thermal reservoir at effective temperature $\beta_\mathrm{h}^\mathrm{eff}$. 
When $\epsilon < 0.5$, we have $(1-\epsilon)/\epsilon > 1$, corresponding to a positive effective temperature ($\beta_\mathrm{h}^\mathrm{eff} > 0$). In this regime, the OIE maps onto a conventional OHE driven by a positive-temperature hot reservoir.
 When $\epsilon > 0.5$, we have $(1-\epsilon)/\epsilon < 1$, yielding a negative effective temperature ($\beta_\mathrm{h}^\mathrm{eff}< 0$). In this regime, the demon induces a state inversion, transforming the engine into one driven by an effective negative-temperature environment.
In this way, we achieved the positive and negative temperature transition of OIE's effective temperature by adjusting the measurement error alone. 
Finally, the transitionless work  $W_{\mathrm{trls}}^{\mathrm{OIE}}$ represents the quasi-static work limit of a OHE driven by this effective temperature, i.e., $W_{\mathrm{trls}}^{\mathrm{OIE}} = W_{\mathrm{qs}}^{\mathrm{OIE}} = W_{\mathrm{qs}}^{\mathrm{OHE}}$.

In the OIE, the energy input to  the system by the demon is $E_\mathrm{MD}$, and the minimum energy required to maintain the demon's continuous operation is $I/\beta_\mathrm{c}$. Therefore,  
the global thermodynamic efficiency of the finite-time quantum OIE is defined as
\begin{equation}\label{effOIE}
\eta_{\mathrm{fin}}^{\mathrm{OIE}} = \frac{W_{\mathrm{fin}}^{\mathrm{OIE}}}{E_{\mathrm{MD}} + I/\beta_\mathrm{c}}.
\end{equation}
\subsection{The OIE can exceed the quasi-static limit}
\subsubsection{$\epsilon\leq0.5$}
We first investigate the operational regime $\epsilon\leq0.5$. In this region, analogous to a conventional finite time quantum OHE coupled to positive-temperature thermal reservoir, quantum internal friction along the adiabatic branches acts to dissipate work.
Because, at $\epsilon = 0.5$, the residual quantum coherence $C(\rho_\mathrm{MD})$ attains its  maximum, while the dissipation associated with the demon vanishes ($I=0$),
we now focus on this optimal case.
The work output then simplifies to
$W_{\mathrm{fin}}^{\mathrm{OIE}} = -(\omega_\mathrm{h} - \omega_\mathrm{c})\langle n_{t_0}\rangle + 2\xi\omega_\mathrm{h}\langle n_{t_0}\rangle + 2\omega_\mathrm{c}\,\zeta_{\mathrm{hc}}^{\mathrm{MD}},$
and the corresponding  efficiency reduces to
$\eta_{\mathrm{fin}}^{\mathrm{OIE}} = W_{\mathrm{fin}}^{\mathrm{OIE}}/E_{\mathrm{MD}}
= \eta_{\mathrm{Otto}} + 2\omega_\mathrm{c}[\xi\langle n_{t_0}\rangle + \zeta_{\mathrm{hc}}^{\mathrm{MD}}]/E_{\mathrm{MD}}.$

Due to $ \xi\langle n_{t_0}\rangle+\zeta_{\mathrm{hc}}^{\mathrm{MD}}\leq\xi\langle n_{t_0}\rangle+|\zeta_{\mathrm{hc}}^{\mathrm{MD}}|
    =\xi\langle n_{t_0}\rangle-2\xi(1-\xi)\langle n_{t_0}\rangle=\xi\langle n_{t_0}\rangle(2\xi-1)>0$ when $\epsilon=0.5$,
 this OIE can extract work by only changing the eigenstates of the Hamiltonian without altering its energy level gap. Specifically, when $\omega_\mathrm{h} = \omega_\mathrm{c}$, the energy level gap of the Hamiltonian remains constant throughout the OIE cycle, while only its eigenstates undergoes modulation.
In this case, the work and efficiency of the OIE reduce to
$ W_{\mathrm{fin}}^{\mathrm{OIE}} = 2\omega_\mathrm{c}(\xi\langle n_{t_0}\rangle + \zeta_{\mathrm{hc}}^{\mathrm{MD}})$
 and $\eta_{\mathrm{fin}}^{\mathrm{OIE}} = 2\omega_\mathrm{c}(\xi\langle n_{t_0}\rangle + \zeta_{\mathrm{hc}}^{\mathrm{MD}})/E_{\mathrm{MD}}$.
Because the maximum value of $\xi\langle n_{t_0}\rangle + \zeta_{\mathrm{hc}}^{\mathrm{MD}}$ is greater than zero, when $\omega_\mathrm{h}=\omega_\mathrm{c}$ the work and efficiency fo the OIE are allowed to be positive, enabling the OIE to produce work  by modulating its phase, as shown in Fig. \ref{e0.5}(a). 
This is fundamentally impossible for a finite-time quantum OHE, as its quasi-static work and efficiency are both zero when $\omega_\mathrm{h} = \omega_\mathrm{c}$.

\begin{figure*}
    \centering
    \includegraphics[width=0.8\linewidth]{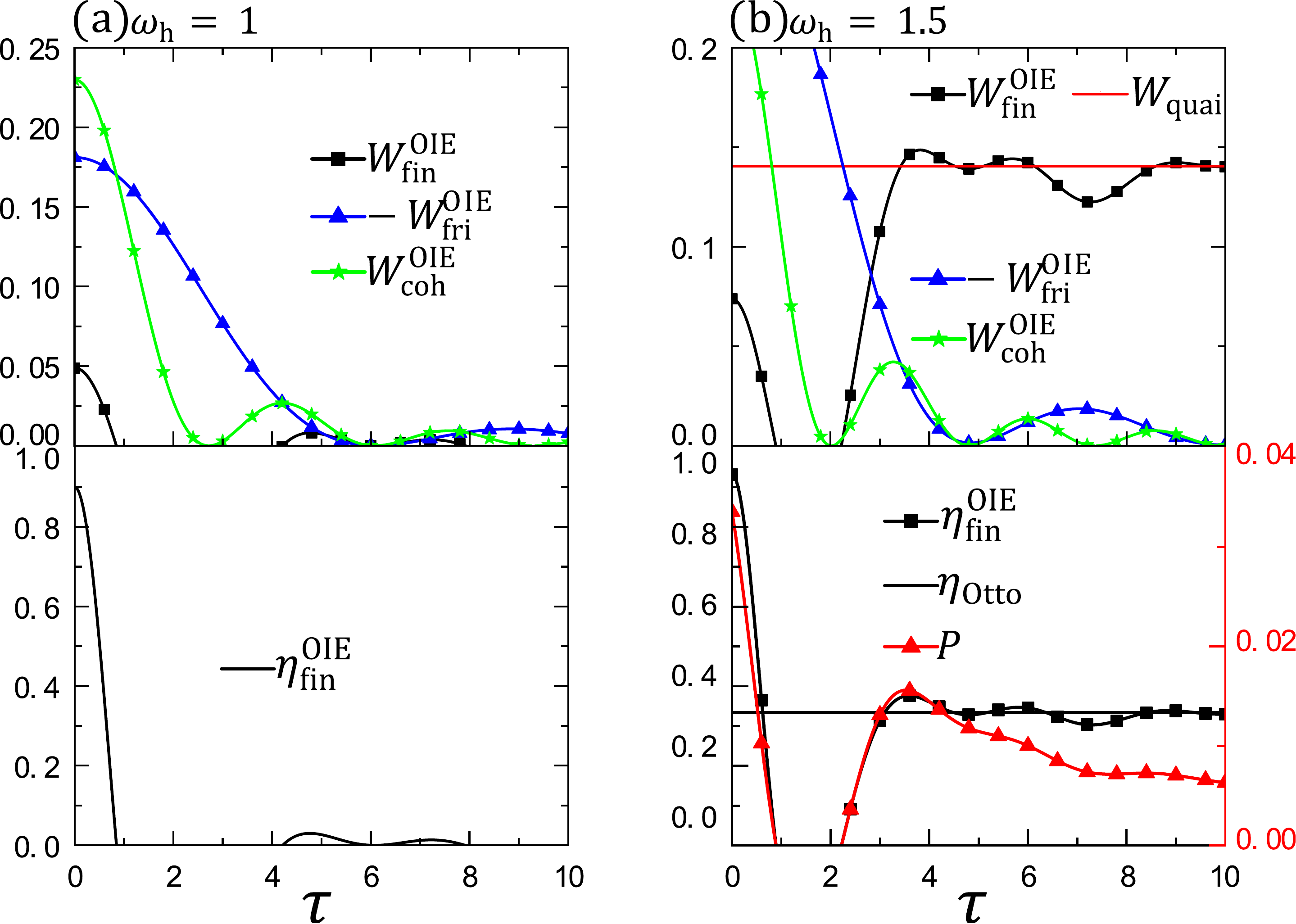}
    \caption{The finite time quantum OIE'  output work and the efficiency   as the functions of the driving time $\tau$ with given $\omega_\mathrm{h}$, under measurement error $\epsilon=0.55$. The right axis of (c) represent the scale of the power. The other parameters are the same as Fig. \ref{Otto}.}
    \label{e0.55}
\end{figure*}

In addition, applying $\xi\langle n_{t_0}\rangle(2\xi-1)>0$ once again, the term $2\xi\omega_\mathrm{h}\langle n_{t_0}\rangle + 2\omega_\mathrm{c}\zeta_{\mathrm{hc}}^{\mathrm{MD}}<2\xi\omega_\mathrm{c}(\langle n_{t_0}\rangle + \zeta_{\mathrm{hc}}^{\mathrm{MD}})$  can also be allowed to exceed zero. This leads directly to the core inequalities
\begin{eqnarray}\label{www}
    W_{\mathrm{fin}}^{\mathrm{OIE}} &>& W_{\mathrm{qs}}^{\mathrm{OIE}}=W_{\mathrm{qs}}^{\mathrm{OHE}} ,\nonumber\\
\eta_{\mathrm{fin}}^{\mathrm{OIE}} &>& \eta_{\mathrm{Otto}}\geq \eta_\mathrm{qs}^\mathrm{OIE},
\end{eqnarray}
where, $\eta_\mathrm{qs}^\mathrm{OIE}=\eta_\mathrm{Otto}/(1+I/\beta_\mathrm{c})$ represents the efficiency of the OIE in the quasi-static limit.
Consequently, as illustrated in Fig. \ref{e0.5}(b), the finite time work output and efficiency of the OIE can simultaneously surpass those of the corresponding OHE in the quasi-static regime. Furthermore, because $W_\mathrm{qs}^\mathrm{OHE}=W_\mathrm{qs}^\mathrm{OIE}$, and $\eta_{\mathrm{Otto}}\geq \eta_\mathrm{qs}^\mathrm{OIE}$, the finite time OIE is capable of exceeding its own quasi-static performance limits. 
This advantage stems from the demon's measurement and feedback can protect quantum coherence. As a result,   the positive work $W_\mathrm{coh}^\mathrm{OIE}$ caused by the  coherence can fully compensate the negative work $W_\mathrm{fri}^\mathrm{OIE}$ induced by quantum internal friction, as shown in Fig. \ref{e0.5}(b).

However, as depicted in Fig. \ref{e0.5}(c), the finite time OIE produces less work than its quasi-static limit. This behavior indicates that the OIE can surpass its quasi-static work bound only when the high frequency $\omega_\mathrm{h}$ is sufficiently close to the low frequency $\omega_\mathrm{c}$, whereas this work advantage gradually fades as $\omega_\mathrm{h} \gg \omega_\mathrm{c}$.
The microscopic physics behind this crossover lies in the frequency weighting of the competing work terms: for a large frequency ratio $\omega_\mathrm{h} / \omega_\mathrm{c}$, the positive work $2\omega_\mathrm{c} \zeta_{\mathrm{hc}}^{\mathrm{MD}}$ derived from quantum coherence is scaled only by the smaller frequency $\omega_\mathrm{c}$, and thus can no longer fully compensate for the  negative work $2\xi \omega_\mathrm{h} \langle n_{t_0}\rangle$ penalty inflicted by internal friction, which grows proportionally with the larger frequency $\omega_\mathrm{h}$.
Intriguingly, the efficiency $\eta_{\mathrm{fin}}^{\mathrm{OIE}}$ can still surpass the quasi-static limit even in the $\omega_\mathrm{h} \gg \omega_\mathrm{c}$ regime. This persistent efficiency boost occurs because the sign of the efficiency enhancement is governed solely by the combination $\xi \langle n_{t_0}\rangle + \zeta_{\mathrm{hc}}^{\mathrm{MD}}$, which remains immune to the large frequency weighting factor $\omega_\mathrm{h}$.

Finally, both the finite time work output $W_{\mathrm{fin}}^{\mathrm{OIE}}$ and the efficiency $\eta_{\mathrm{fin}}^{\mathrm{OIE}}$ are  continuous functions of the measurement error $\epsilon$. Consequently, all the above quantum advantages persist for imprecise measurements within a finite neighborhood sufficiently close to the optimal point $\epsilon = 0.5$. For instance, as explicitly demonstrated in Appendix \ref{saw}, at $\epsilon = 0.45$, the finite time OIE retains its ability to extract net work solely through phase modulation, while simultaneously outperforming the corresponding quasi-static OHE in both work output and efficiency.

\subsubsection{$\epsilon\geq0.5$}

In contrast, for the  regime where $\epsilon > 0.5$ ($\langle n_{t_2}^{\mathrm{MD}}\rangle > 0$), the engine operates in a population-inverted state. Here, the synergistic interplay between the quantum coherence work $2\omega_\mathrm{c} \zeta_{\mathrm{hc}}^{\mathrm{MD}}$ and the positive quantum friction work $2\xi \omega_\mathrm{c} \langle n_{t_2}^{\mathrm{MD}}\rangle$ generated during the adiabatic expansion stroke  offsets the negative quantum friction work $2\xi \omega_\mathrm{h} \langle n_{t_0}\rangle$ arising from adiabatic compression. This combined mechanism enables the finite-time OIE to surpass its quasi-static benchmarks.

Because the mutual information is strictly bounded above by $I \le \ln 2$, the direct feedback energy $E_{\mathrm{MD}}$ injected by the demon can be significantly larger than the demon's dissipation $I/\beta_\mathrm{c}$. Under this dominant feedback energy condition, the  finite time efficiency of the OIE simplifies to
$\eta_\mathrm{fin}^\mathrm{OIE}=\eta_\mathrm{Otto}+2\omega_\mathrm{c}[\xi(\langle n_{t_0}\rangle +\langle n_{t_2}^\mathrm{MD}\rangle)+\zeta_\mathrm{hc}^\mathrm{MD}]/E_\mathrm{MD}.$
Consequently, whenever the second term on the right-hand side remains positive in the $\epsilon > 0.5$ regime, the efficiency of the OIE under this condition is allowed to surpass its corresponding quasi-static Otto efficiency. 

According to the analytical expression for the finite-time OIE' work output, the work $W_{\mathrm{fin}}^{\mathrm{OIE}}$ exceeds its quasi-static limit whenever the constructive coherence term $\zeta_{\mathrm{hc}}^{\mathrm{MD}}>0$ alongside  the average quantum number $\langle n_{t_2}^{\mathrm{MD}}\rangle$ satisfies $\langle n_{t_2}^{\mathrm{MD}}\rangle > -(\omega_\mathrm{h} / \omega_\mathrm{c})\langle n_{t_0}\rangle$. Under this inequality,  naturally the relation $\xi(\langle n_{t_0}\rangle +\langle n_{t_2}^\mathrm{MD}\rangle)+\zeta_\mathrm{hc}^\mathrm{MD}>0$ is satisfied, guaranteeing that the efficiency $\eta_\mathrm{fin}^\mathrm{OIE}$ surpasses $\eta_{\mathrm{Otto}}$.
 From the inequality $\langle n_{t_2}^\mathrm{MD}\rangle>-\omega_\mathrm{h}/\omega_\mathrm{c}\langle n_{t_0}\rangle$, we obtain $(2\epsilon - 1)/(2\rho_{t_0}^{gg} - 1) > \omega_\mathrm{h}/\omega_\mathrm{c}$. Consequently, for any measurement error $\epsilon>\rho_{t_0}^{gg}$, this condition can be readily satisfied by selecting an appropriate driving frequency ratio  $\omega_\mathrm{h}/\omega_\mathrm{c}$.  In fact, because the  coherence work $2\omega_\mathrm{c} \zeta_{\mathrm{hc}}^{\mathrm{MD}}$ provides additional positive work to compensate for adiabatic compression  losses, the physically accessible range of $\omega_\mathrm{h}$ is  not strictly bounded by $(2\epsilon - 1) / [\omega_\mathrm{c}(2\rho_{t_0}^{gg} - 1)]$. 
 
Thus, for $\epsilon>0.5$,
due to the quantum internal friction during adiabatic expansion acts as a work generator rather than a loss, working hand-in-hand with quantum coherence to overcome compression friction,  both the work output and efficiency of the finite-time  quantum OIE can simultaneously exceed those of the corresponding quasi-static quantum OHE, ultimately breaking its own quasi-static thermodynamic bounds, as illustrated in Fig. \ref{e0.55}. 

Furthermore, as illustrated in Fig. \ref{e0.55}, in the fast-driving regime $\tau\to 0$,  the quantum OIE not only yields a positive  work output, but this work output monotonically increases as the driving time  shortens. This behavior arises that  a shorter driving time enhances the transition probability  $\xi$ along both adiabatic strokes, while the  work $2\omega_\mathrm{c}\xi\langle n_{t_2}^\mathrm{MD}\rangle+2\omega_c \zeta_\mathrm{hc}^\mathrm{MD}$ increases faster than the negative work $2\omega_\mathrm{h}\xi\langle n_{t_2}^\mathrm{MD}\rangle$ decreases as $\xi$ increases. This can be achieved, for example, when relation $\langle n_{t_2}^{\mathrm{MD}}\rangle > -(\omega_\mathrm{h} / \omega_\mathrm{c})\langle n_{t_0}\rangle$ is satisfied.
 In contrast, for the regime where $\epsilon\leq0.5$, such behavior is impossible. In that case, as $\xi$ grows with faster driving, the negative work generated by the internal friction across both adiabatic strokes escalates faster than the corresponding growth rate of the positive coherence work, ultimately suppressing work extraction.
Finally, owing to this  fast-driving feature, the output power $P=W_\mathrm{fin}^\mathrm{OIE}/\tau_\mathrm{cyc}$ of the finite time OIE in the $\epsilon > 0.5$ regime increases continuously as the driving time decreases.

On the other hand, according to Eq. (\ref{Emd}), the energy injected into the system by the Maxwell's demon, $E_{\text{MD}}$, decreases as the transition probability $\xi$ increases, whereas the work output grows with increasing $\xi$ in the fast-driving limit ($\tau \to 0$). Consequently, as depicted in Fig. \ref{e0.55}, the efficiency of the OIE increases dramatically as the driving duration $\tau$ is shortened. As a result, the OIE can simultaneously achieve maximum power output and maximum efficiency within this rapid-driving regime, fundamentally overcoming the classic trade-off between power and efficiency inherent to conventional thermal engines. This provides a viable blueprint for designing next-generation, high-power and high-efficiency micro/nanoscale quantum machines.

\subsection{The finite time OIE  can achieve almost perfect efficiency}

According to Eq. (\ref{effOIE}), to approach perfect efficiency  ($\eta_{\mathrm{fin}}^{\mathrm{OIE}} \to 1$) in the OIE, a fundamental thermodynamic prerequisite is that the work output $W_{\mathrm{fin}}^{\mathrm{OIE}}$ approaches the total energy cost, $E_{\mathrm{MD}} + I/\beta_\mathrm{c}$. To fulfill this condition, two requirements must be satisfied simultaneously: the extracted mutual information $I$ must approach zero, and  $W_{\mathrm{fin}}^{\mathrm{OIE}}$ must approach $E_{\mathrm{MD}}$. The former requirement is readily satisfied by tuning the measurement error $\epsilon$ toward the weak-measurement limit $\epsilon = 0.5$. The latter condition however, necessitates that the heat dissipated by the OIE into the cold reservoir vanishes. In the following, we explicitly analyze this zero-dissipation scenario under $\epsilon=0.5$.

To test whether zero heat dissipation is physically permissible, we set the energy released to the cold reservoir  to zero, i.e., $E_{t_0} = E_{t_3}^{\mathrm{MD}}$, which yields the constraint on the average quantum numbers and coherence
$ \langle n_{t_0}\rangle=(1-2\xi)\langle n_{t_2}^\mathrm{MD}\rangle-2\zeta_\mathrm{hc}^\mathrm{MD}$.
We first analyze the operational regime of non-sudden driving where $\xi<0.5$. Since $E_\mathrm{MD}>0$,  the system must satisfy the relation $\langle n_{t_2}^\mathrm{MD}\rangle>(1-2\xi) \langle n_{t_0}\rangle$, indicating
$\langle n_{t_0}\rangle>(1-2\xi)^2\langle n_{t_0}\rangle-2\zeta_\mathrm{hc}^\mathrm{MD}.$
For this inequality to hold, the left-hand side must exceed the minimum value of the right-hand side. The right-hand side attains its minimum when $\zeta_\mathrm{hc}^\mathrm{MD}$  reaches its maximum value $\zeta_\mathrm{hc}^\mathrm{MD}=-2(1-\xi)\xi\langle n_{t_0}\rangle$.
Substituting this  maximum value into  the previous inequality transforms it into
$\langle n_{t_0}\rangle>(1-2\xi)^2\langle n_{t_0}\rangle+4(1-\xi)\xi\langle n_{t_0}\rangle $.
Dividing both sides by  $\langle n_{t_0}\rangle$ simplifies the inequality to
 $(1-2\xi)^2+4(1-\xi)\xi-1>0$. However, direct algebraic expansion reveals that $(1 - 2\xi)^2 + 4(1 - \xi)\xi - 1 = 0$. 
 Thus, achieving zero heat dissipation into the cold reservoir is  impossible for $\xi<0.5$.
 
\begin{figure}
    \centering
    \includegraphics[width=1\linewidth]{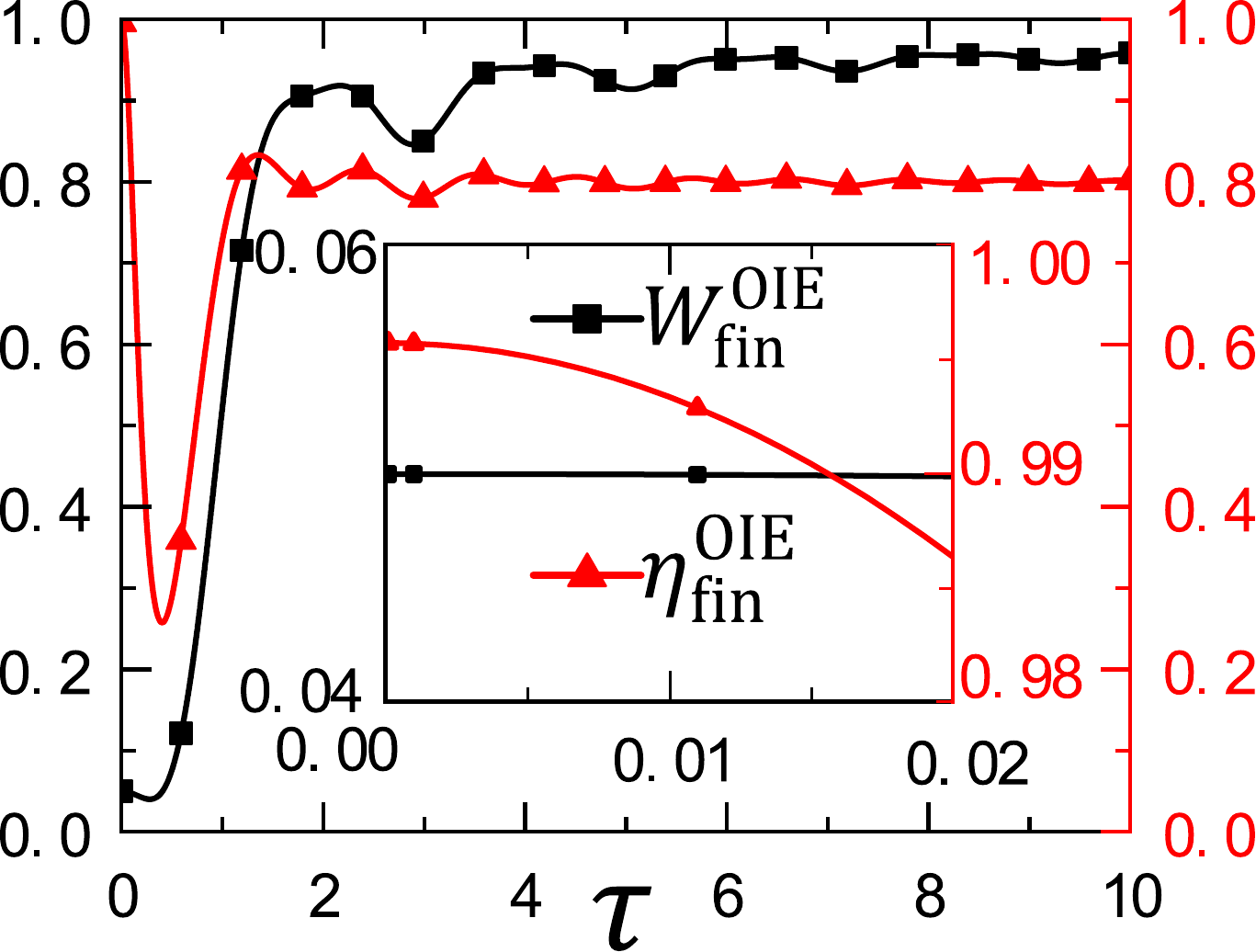}
    \caption{The output work and the efficiency of the finite time quantum OIE as functions of the driving time $\tau$ with  measurement error $\epsilon=0.51$. The other parameters are the same as Fig. \ref{Otto}.}
    \label{e0.51}
\end{figure}

 Next, we examine the critical limit where $\xi=0.5$, In this limit,  the condition for zero heat dissipation is  $\langle n_{t_0}\rangle=-2\zeta_\mathrm{hc}^{MD}$.
 Since the maximum possible value of the coherence term $\zeta_{\mathrm{hc}}^{\mathrm{MD}}$, given by $-2(1-\xi)\xi\langle n_{t_0}\rangle$, evaluates precisely to $-\langle n_{t_0}\rangle / 2$ when $\xi = 0.5$, the  equality $\langle n_{t_0}\rangle=-2\zeta_\mathrm{hc}^{MD}$ can be satisfied. Furthermore, for $\xi = 0.5$, the internal energy after adiabatic compression vanishes, $E_{t_1} = (1 - 2\xi)\langle n_{t_0}\rangle = 0$, which simplifies the work output  to $W_{\mathrm{fin}}^{\mathrm{OIE}} = E_{\mathrm{MD}} = \omega_\mathrm{h} (\epsilon - 0.5)$.
In this configuration, if $\epsilon=0.5-\delta$ (where $\delta$ is a small positive quantity), we obtain $W_\mathrm{fin}^\mathrm{OIE}<0$ which lacks physical significance for an engine mode. Conversely, if $\epsilon=0.5+\delta$ the work output satisfies $W_\mathrm{fin}^\mathrm{OIE}>0$.  Consequently, as $\tau\to 0$ ($\xi\to0.5$) and the measurement error  $\epsilon$ approaches $0.5$ from above  ($\epsilon\to0.5^+$), the finite-time OIE achieves near-perfect efficiency while maintaining positive work extraction, as shown in Fig. \ref{e0.51}.

As illustrated in this figure, when $\tau \to 0$, the thermodynamic efficiency of the finite-time OIE exceeds $0.99$, virtually attaining unit efficiency. In addition, during the cold isochoric stroke, the non-equilibrium factor $e^{-\Gamma_\mathrm{c} \tau_\mathrm{c}/2}$, depends directly on the coupling parameter $\gamma_0$ and the stroke duration $\tau_\mathrm{c}$. Therefore, one can enhance $\gamma_0$ by increasing the photon spectral density at resonant frequency $\omega_\mathrm{c}$, thereby rendering the system's thermal relaxation time far shorter than $\tau$. Furthermore, in the limit $\tau \to 0$, because the work output of the OIE grows as $\tau$ decreases, the power output shoots up rapidly with decreasing $\tau$. Under these conditions, the OIE simultaneously achieves an exceptionally high power output and a near-unity efficiency.

\section{Summary}
 We first demonstrate that for a conventional finite-time quantum Otto heat engine coupled to two thermal Gibbs reservoirs, both the work output and thermodynamic efficiency are strictly bounded from above by their quasi-static counterparts.
 We then demonstrate that when a Maxwell's demon replaces the hot reservoir to power the engine, called the quantum Otto information engine (OIE), the engine can extract work by adjusting the eigenstates of Hamiltonian  instead of adjusting the energy gap.  Even when the energy gap is modulated, both the work output and the efficiency (which contains the demon' dissipation) of the OIE  can exceed the quasi-static bounds of the corresponding Otto heat engine.
This enhancement stems from the demon’s ability to preserve quantum coherence while effectively converting quantum internal friction into useful work.
Remarkably, in the rapid-driving and critical-measurement regime ($\tau \to 0, \epsilon \to 0.5^+$), the OIE achieves near-unity efficiency while sustaining positive work extraction.
Given  the finite time Otto cycle \cite{Lutz16,Lutz24} and the Maxwell’s demon \cite{Yan24,Yan26} have been individually demonstrated using $^{40}\mathrm{Ca}^+$, our proposed OIE is readily accessible with present experimental capabilities. Ultimately, our results can pave the way for practical quantum technologies, offering compelling avenues for high power and high efficiency  Micro/nano machines and on-chip thermal management due to produce work by the  modulation of the Hamiltonian phase.
\begin{acknowledgments}
 Y. Xiao thanks the support in part by the National Natural Science Foundation of China Grant No. NSFC 12234019.  
\end{acknowledgments}
\nocite{*}

\appendix
\begin{widetext}
\section{Density matrix elements of the system at $t_3$ time}\label{app1}
\begin{eqnarray}
\rho_{t_3}^{ee} &=& \langle e_{t_0}|\rho_{t_3}|e_{t_0}\rangle
= \langle e_{t_0}|U_{\mathrm{hc}}\rho_{t_2}U_{\mathrm{hc}}^\dagger|e_{t_0}\rangle= \rho_{t_2}^{ee}\langle e_{t_0}|U_{\mathrm{hc}}|e_{t_1}\rangle\langle e_{t_1}|U_{\mathrm{hc}}^\dagger|e_{t_0}\rangle\nonumber\\
&+&\rho_{t_2}^{eg}\langle e_{t_0}|U_{\mathrm{hc}}|e_{t_1}\rangle\langle g_{t_1}|U_{\mathrm{hc}}^\dagger|e_{t_0}\rangle+\rho_{t_2}^{ge}\langle e_{t_0}|U_{\mathrm{hc}}|g_{t_1}\rangle\langle e_{t_1}|U_{\mathrm{hc}}^\dagger|e_{t_0}\rangle\nonumber\\
 &+&\rho_{t_2}^{gg}\langle e_{t_0}|U_{\mathrm{hc}}|g_{t_1}\rangle\langle g_{t_1}|U_{\mathrm{hc}}^\dagger|e_{t_0}\rangle=\rho_{t_2}^{ee}(1-\xi) + \rho_{t_2}^{gg}\xi+2\mathrm{Re}(\rho_{t_2}^{ge}U_\mathrm{hc}^{eg}U_\mathrm{hc}^{ee*})\nonumber\\
&=&\rho_{t_2}^{ee}(1-\xi) + \rho_{t_2}^{gg}\xi-2\mathrm{Re}(\rho_{t_2}^{ge}U_{hc}^{gg}U_{hc}^{ge*}),\\
\rho_{t_3}^{eg} &=& \langle e_{t_0}|\rho_{t_3}|g_{t_0}\rangle
= \langle e_{t_0}|U_{\mathrm{hc}}\rho_{t_2}U_{\mathrm{hc}}^\dagger|g_{t_0}\rangle = \rho_{t_2}^{ee}\langle e_{t_2}|U_{\mathrm{hc}}|e_{t_1}\rangle\langle e_{t_1}|U_{\mathrm{hc}}^\dagger|g_{t_0}\rangle\nonumber\\
&+&\rho_{t_2}^{eg}\langle e_{t_0}|U_{\mathrm{hc}}|e_{t_1}\rangle\langle g_{t_1}|U_{\mathrm{hc}}^\dagger|g_{t_0}\rangle+rho_{t_2}^{ge}\langle e_{t_0}|U_{\mathrm{hc}}|g_{t_1}\rangle\langle e_{t_1}|U_{\mathrm{hc}}^\dagger|g_{t_0}\rangle\nonumber\\
 &+&\rho_{t_2}^{gg}\langle e_{t_0}|U_{\mathrm{hc}}|g_{t_1}\rangle\langle g_{t_1}|U_{\mathrm{hc}}^\dagger|g_{t_0}\rangle=2\langle n_{t_2}\rangle+\rho_{t_2}^{eg}U_\mathrm{hc}^{ee}U_\mathrm{hc}^{gg*}+\rho_{t_2}^{ge}U_\mathrm{hc}^{eg}U_\mathrm{hc}^{ge*},\\
\rho_{t_3}^{gg} &=& \langle g_{t_0}|\rho_{t_3}|g_{t_0}\rangle
= \langle g_{t_0}|U_{\mathrm{hc}}\rho_{t_2}U_{\mathrm{hc}}^\dagger|g_{t_0}\rangle = \rho_{t_2}^{ee}\langle g_{t_0}|U_{\mathrm{hc}}|e_{t_1}\rangle\langle e_{t_1}|U_{\mathrm{hc}}^\dagger|g_{t_0}\rangle\nonumber\\
&+&\rho_{t_2}^{eg}\langle g_{t_0}|U_{\mathrm{hc}}|e_{t_1}\rangle\langle g_{t_1}|U_{\mathrm{hc}}^\dagger|g_{t_0}\rangle+\rho_{t_2}^{ge}\langle g_{t_0}|U_{\mathrm{hc}}|g_{t_1}\rangle\langle e_{t_1}|U_{\mathrm{hc}}^\dagger|g_{t_0}\rangle\nonumber\\
 &+&\rho_{t_2}^{gg}\langle g_{t_0}|U_{\mathrm{hc}}|g_{t_1}\rangle\langle g_{t_1}|U_{\mathrm{hc}}^\dagger|g_{t_0}\rangle=\rho_{t_2}^{ee}\xi+\rho_{t_2}^{gg}(1-\xi)+2\mathrm{Re}(\rho_{t_2}^{ge}U_{hc}^{gg}U_{hc}^{ge*}).
\end{eqnarray}

Here, we have used the relation $\rho_{t_2}^{ge}U_{\mathrm{hc}}^{eg}U_{\mathrm{hc}}^{ee*} = -\rho_{t_2}^{ge}U_{\mathrm{hc}}^{gg}U_{\mathrm{hc}}^{ge*}$.

\section{The performance of the OIE at $\epsilon=0.45$}\label{saw}

Here, we present the work output and thermodynamic efficiency of the Otto information engine as functions of the driving time at $\epsilon = 0.45$. Given the consistent dynamical behavior observed between Figs. \ref{figs} and \ref{e0.5}, our conclusions remain valid for imprecise measurements within a finite neighborhood sufficiently close to $\epsilon = 0.5$. 
\begin{figure*}
    \centering
\includegraphics[width=1\linewidth]{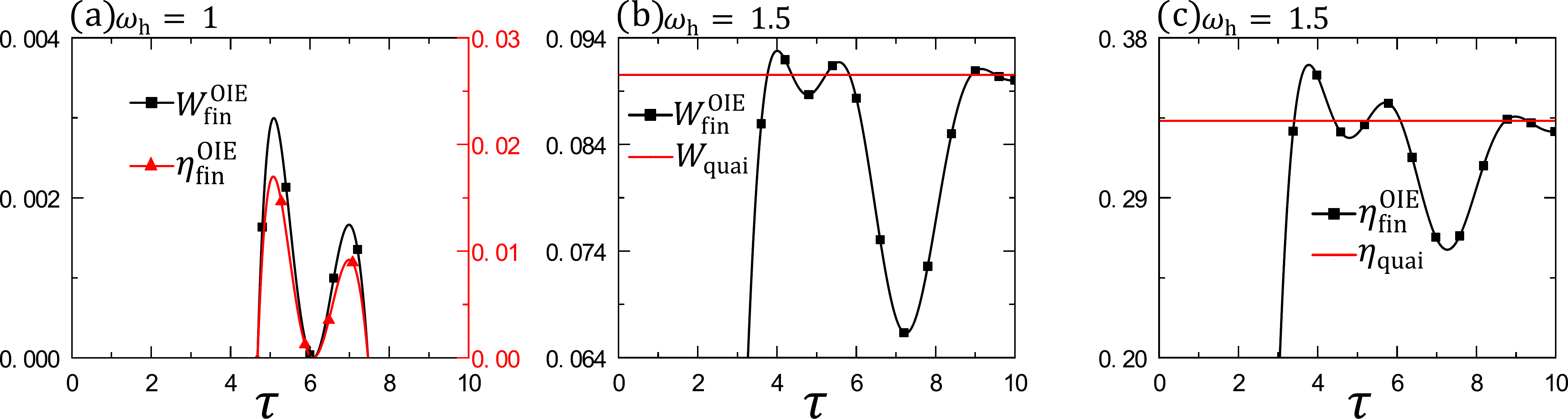}
    \caption{The output work and the efficiency of the finite time quantum OIE as functions of the driving time $\tau$ with given $\omega_\mathrm{h}$ under measurement error $\epsilon=0.45$. The other parameters are the same as Fig. \ref{Otto}.}
    \label{figs}
\end{figure*}
\end{widetext}
\bibliography{ref}

\end{document}